\documentclass[superscriptaddress,aps,prl,preprintnumbers,twocolumn]{revtex4}

\usepackage{epsfig,amsmath,amssymb}
\usepackage[latin1]{inputenc}

\renewcommand{\epsilon}{\varepsilon}
\providecommand{\url}[1]{\texttt{#1}}

\begin{document}

\title{Mapping out of equilibrium into equilibrium: the macroscopic 
fluctuations
of simple transport models}


\author{Julien Tailleur}
\affiliation{PMMH-ESPCI, CNRS UMR 7636, 10 rue Vauquelin, 75005
  Paris, France}

\author{Jorge Kurchan}
\affiliation{PMMH-ESPCI, CNRS UMR 7636, 10 rue Vauquelin, 75005
  Paris, France}

\author{Vivien Lecomte}
\affiliation{MSC, CNRS UMR 7057, Universit\'e Paris VII,
        10 rue Alice Domon et L\'eonie Duquet, 75013 Paris, France}
\date{\today}

\begin{abstract}
  We study a simple transport model driven out of equilibrium by
  reservoirs at the boundaries, corresponding to the hydrodynamic
  limit of the Symmetric Simple Exclusion Process (SSEP). We show that
  a non-local transformation of densities and currents maps the large
  deviations of the model into those of an open, isolated chain
  satisfying detailed balance, where rare fluctuations are the
  time-reversals of relaxations. We argue that the existence of such a
  mapping is the immediate reason why it is possible for this model to
  obtain an explicit solution for the large-deviation function of
  densities through elementary changes of variables. This approach can
  be generalized to the other models previously treated with the
  Macroscopic Fluctuation Theory \cite{Bertini01}.
\end{abstract}


\maketitle

Systems driven by reservoirs at the borders are possibly the first
step in complication as one enters the non-equilibrium realm.  They
are already not solvable in general, but many elegant and striking
results have been obtained in recent years (for a review, see
\cite{Derrida07} and references therein). Recently, Derrida, Lebowitz
and Speer~\cite{Derrida01} (DLS) obtained an exact expression for the
large deviation function of the density profile in the Symmetric
Simple Exclusion Model (SSEP) -- a one-dimensional transport of
particles -- using a matrix method that had been developed
previously~\cite{Derrida94}. This is a major achievement, as large
deviation functions are a natural extension of free energy to out of
equilibrium situations.

For the sake of generality, it is desirable to have a purely
macroscopic approach, that does not rely on the strong symmetries of
the microscopic model. This was done within the Macroscopic
Fluctuation Theory~\cite{Bertini01,Bertini05}. This is a version of
WKB theory, valid in the limit of large coarse-graining scale $N$: in
the usual manner of Semiclassical Theory and Geometric Optics, the
logarithm of wavefunctions evolving with the Fokker Planck operator
obey a Hamilton-Jacobi equation whose characteristics are
trajectories satisfying Hamilton's equations.

For the hydrodynamic limit of the SSEP, Bertini et al.\cite{Bertini01}
were able to integrate explicitly the corresponding Hamilton-Jacobi
equations and recovered the large-deviation function. They thus showed
that such a solution is in principle logically independent of the
exact microscopic solution. Their derivation amounts to a series of
carefully chosen changes of variables, each of one yielding a miracle,
whose cause and degree of generality one may still wish to ascertain.

In this paper we uncover a transformation of the large deviations of
densities and currents of the original driven model into those of a
'dual' isolated, equilibrium chain. Large deviations and optimal
trajectories are easily obtained in this representation using the
detailed balance property, and can then be mapped back to the original
setting. Moreover, emergences of rare fluctuations in the dual model
are the time reverse of relaxations to the average profile, but this
symmetry is lost in the mapping back to the original model. This
accounts in this case for the lack of Onsager-Machlup
symmetry~\cite{Onsager} between birth and death of a fluctuation,
which has received considerable interest~\cite{Luchinsky,Ciliberto,Bertini01}
over the past few years.

In what follows, we shall first rephrase the Macroscopic Fluctuation
Theory in a path integral language, then single out the specific
properties which enable one to compute the large deviation function
for the system (\ref{eqn:hydro}) and finally encapsulate them in the
existence of a dual model with detailed balance.
\begin{center}
  {\bf Hydrodynamic Model}
\end{center}
We consider the fluctuating hydrodynamic limit of the
SSEP in contact with two reservoirs~\cite{Spohn83}
\begin{equation}
  \begin{aligned}
      \label{eqn:hydro}
      &\dot \rho = -\nabla J;
      \qquad \;\;\,J = - \frac{1}{2} \nabla \rho - \sqrt{\sigma_\rho} \eta \\
      &\rho(0,t)=\rho_0;\qquad \rho(L,t)=\rho_L
    \end{aligned}
\end{equation}
Here $\eta$ is a white
noise of variance $1/N$ and $ \sigma_\rho=\rho (1-\rho)$.
The probability of a trajectory is given by~\cite{Zinn}
\begin{equation*}
  \begin{aligned}
    P &\sim \int {\cal D}[\rho,\eta] \delta \left(\dot \rho +
      \nabla J\right)e^{-N \int \text{d}t \text{d}x \,\frac {\eta^2}2}\\
    &\sim \int {\cal D}[\rho, \hat \rho,\eta] e^{-N \int\text{d}t
      \text{d}x \left\{\hat \rho \left[\dot \rho - \frac{1}{2}
          \Delta \rho - \nabla( \sqrt{ \sigma_\rho} \eta) \right] +
        \frac {\eta^2}2\right\}}\\
  \end{aligned}
\end{equation*}
After an integration by parts (which shall be frequent and unannounced
in the following), the integration over $\eta$ gives
\begin{equation}
  \label{eqn:action}
  \int {\cal D}[\hat \rho,\rho] e^{-N S[\hat \rho, \rho]}=\int
  {\cal D}[\hat \rho,\rho]e^{- N \int\text{d}t \text{d}x \{ \hat
    \rho \dot \rho - {\cal {H}}\}}
\end{equation}
where we have introduced the Hamiltonian density defined by $H=\int\!\!
{\cal H}\,\text d x$:
\begin{equation}
\label{eqn:hamdens}
{\cal{H}}\equiv \frac{1}{2} \left[ \sigma_\rho \nabla \hat \rho^2
+ \hat \rho \Delta \rho\right]
\end{equation}

\begin{center}
{\bf Large Deviations: General strategy}
\end{center}
In the large-$N$ limit, the probability of observing a profile
$\rho^*(x)$ in the steady state scales as $P(\rho^*) \sim e^{-N {\cal
    F}[\rho^*(x)]}$.  We wish to calculate the large-deviation
function ${\cal F}$, which is given by the action of the 'instanton'.
This is the trajectory starting in a neighbourhood of the stationary
profile $\bar \rho$ that extremises the action (\ref{eqn:action}),
converges to $\rho^*(x)$ at a large time $T$ (and thus has at all
times $H=0$) and satisfies the spatial
boundary conditions. The problem then reduces to solving the equations
of motion:
\begin{equation}
\label{eqn:hameq}
\left\{
  \begin{aligned}
    \dot \rho &= \frac{\delta {H}}{\delta \hat \rho(x)} =
\frac 1 2 \Delta \rho - \nabla [ \sigma_\rho \nabla \hat \rho]\\
    \dot {\hat \rho} &=- \frac{\delta {H}}{\delta  \rho(x)}= 
-\frac 1 2 \Delta \hat \rho + (2\rho-1) \frac{\nabla
      \hat \rho^2}2
  \end{aligned}
  \right.
\end{equation}
with the space and time constraints
\begin{equation}
\label{eqn:BC}
  \begin{aligned}
    \rho(x,0)&=\bar \rho(x)=\rho_0+x 
\frac {\rho_L-\rho_0}L;\quad \rho(x,T)=\rho^*(x)\\
    \rho(0,t)&=\rho_0;\quad
    \rho(L,t)=\rho_L;\quad  \hat \rho(0,t)= \hat \rho(L,t)=0
  \end{aligned}
\end{equation}
The last equality simply says that no fluctuations are allowed at the
contact with the reservoir~\cite{Spohn83,Bertini01}. Alternatively, one can
solve the classical equations (\ref{eqn:hameq}) via the
Hamilton-Jacobi formalism~\cite{Goldstein} and that amounts to the
strategy followed by Bertini et al.
\begin{center}
  {\bf Mapping the problem into a  downhill one. Detailed balance}
\end{center}
The fact that equations (\ref{eqn:hameq}) derive from a stochastic
problem implies that there is a family of explicit `downhill' (zero
noise) solutions: $\hat \rho(t)=\text{C}^{st}$ and $\dot \rho = \frac
1 2 \Delta \rho$. For $\text{C}^{st}=0$, the corresponding action
$S[\rho,\hat \rho]$ is zero. Such a solution is not what we are
looking for, as it relaxes into and not out of the stationary state
and hence does not satisfy the boundary conditions in
time~(\ref{eqn:BC}).

A strategy to find solutions of equations like (\ref{eqn:hameq}) is to
make a change of variables that maps the original problem into another
one of the same form -- i.e. that formally derives from some {\em
  other} stochastic problem, but such that the downhill solutions of
the new problem obey the correct boundary conditions in time and in
space~\footnote{Conversely, if one knows the large deviation function
  ${\cal F(\rho)}$, such a transformation is given by $\hat \rho \to
  \hat \rho + \frac{\delta {\cal F}}{\delta \rho }$~\cite{Bertini01}.}.

For a chain at equilibrium, a simple procedure can be followed to do
so, taking advantage of the detail balance relation of Hamiltonian
(\ref{eqn:hamdens}), which can be made explicit by writing
\begin{equation*}
  {\cal H}=\frac 1 2 \nabla \hat \rho \; \sigma_\rho \; \nabla \left[\hat \rho
      - \log \frac{\rho}{1- \rho} \right]=\frac 1 2
    \nabla \hat \rho \; \sigma_\rho \; \nabla \left[\hat \rho
      - \frac{\delta V_\rho}{\delta \rho} \right]
\end{equation*}
with $ V_\rho = \int \text{d} x [\rho \log \rho + (1-\rho)
\log(1-\rho)]$.
At the level of the action~\footnote{In this paper we shall only invoke
detailed balance at  the large $N$
  'classical' level}, detailed balance means that this form is
left invariant by a succession of two transformations
\begin{equation}
  \label{eqn:shift}
  \hat \rho \to \hat \rho + \log
  \frac \rho {1-\rho}=\hat \rho + \frac{\delta V_\rho}{\delta \rho};\quad (\hat \rho,t) \to (-\hat \rho, T-t)
\end{equation}
The first shift maps
\begin{equation}
\label{ppp}
  \begin{aligned}  
    {\cal H}&\to \tilde {\cal H}=\frac 1 2 \nabla \hat \rho \; \sigma_\rho \;
    \nabla \left[\hat \rho + \frac{\delta V_\rho}{\delta \rho} \right]\\
    S[\hat \rho, \rho] &\to [V_\rho]_0^T +  \int \text{d}
 t \text{d}x \left\{ \hat \rho \dot \rho - \tilde {\cal H}\right\}
  \end{aligned}
\end{equation}
The new $\tilde {\cal H}$ has the form of a stochastic problem, and
the equations of motion associated to (\ref{ppp}) admit a 'downhill'
solution $\hat \rho=\text{C}^{st}$ which in the old variables reads
\begin{equation}
  \label{eqn:uphill}
  \dot \rho = -\frac 1 2 \Delta \rho
  \qquad  \hat \rho = \log \frac \rho {1-\rho}+\text{C}^{st}
\end{equation}
and corresponds to the optimal uphill trajectory. The second shift in
(\ref{eqn:shift}) shows that this trajectory is the time reverse of a
diffusive trajectory. For a chain driven out of equilibrium by the
boundaries, this simple strategy fails: (\ref{eqn:uphill}) is not
compatible with the spatial boundary conditions (\ref{eqn:BC}) for
$\hat \rho$.

Before going on, let us note a striking property, specific to the
problem (\ref{eqn:hydro}).  Rearranging the Hamiltonian density
\begin{equation}
\label{eqn:NLDB}
\!\!{\cal H}\!\!=\!- \rho{\frac{(\nabla \hat \rho)}2}^{\!2}\!\!  \left[ \rho-\!1\!
    - \!\frac{\Delta\hat  \rho}{(\nabla \hat \rho)^2} \right]
  \!\!=\! - \rho {\frac{ (\nabla \hat \rho)}2}^{\!2} \!\! \left[ \rho -\!1\!
    - \! \frac{\delta V_{\hat \rho}}{\delta \hat \rho} \right]\!\!\!\!\!\!\!\!\!
\end{equation}
where $V_{\hat \rho} = \int \text{d} x [ \log \nabla \hat
\rho]$, ${\cal H}$ can {\em formally} now be seen as deriving from
another stochastic dynamics:
$ \dot {\hat \rho}=-\frac 1 2 \Delta \hat \rho -\frac 1 2 {(\nabla
  \hat \rho)} ^2+ \eta \nabla \hat \rho$ with a further detailed
balance symmetry, induced by the two transformations
\begin{equation}
\label{eqn:BrokNLDB}
  \rho \to \rho +\frac{\delta V_{\hat \rho}}{\delta \hat \rho};\qquad (\rho,t)\to
(1-\rho,T-t)
\end{equation}
As in the previous paragraph, two further classes of solutions can be
directly read in (\ref{eqn:NLDB}): $\rho=0$, $\dot {\hat \rho}=-\frac
1 2 \Delta \hat \rho-\frac 1 2 \nabla \hat \rho^2$ and
$\rho-1=\frac{\delta V_{\hat \rho}}{\delta \hat \rho}$, $\dot {\hat
  \rho}=\frac 1 2 \Delta \hat \rho+\frac 1 2 \nabla \hat \rho^2$,
respectively. However, none of these solutions satisfies the boundary
conditions. This additional symmetry is a signature of the existence
of the dual model, as we shall see below. This is the first time the
specific form of the model (in particular its one dimensional nature)
plays a role.

\begin{center}
{\bf A Solution for the classical problem}
\end{center}
Let us now paraphrase Bertini et al., following a concise but
unrigorous Hamiltonian -- rather than Hamilton-Jacobi -- approach.  To make
contact with the exact solution, we rewrite the intermediate variable
$\hat \rho$ appearing in equation (\ref{ppp}) in terms of the DLS
variable $F$ defined by $F=(1+e^{\hat \rho})^{-1}$. To keep the action
in a Hamiltonian form, we also introduce the canonically conjugate
variable $\hat F=\frac{\rho}{F(1-F)}$. This maps the action into
\begin{equation*}
  \begin{aligned}
    S&=\int \text{d}x \left[\rho \log \rho + (1-\rho) \log(1-\rho)+\rho 
\log \frac{1-F}F\right]_0^T\\
    &+  \int \text{d}x
    \text{d}t \left\{ \hat F \dot F +\frac 1 2 \hat F \nabla F^2\left[\hat F -
        \frac 2 {1-F}\right] +\frac 1 2 \nabla \hat F \nabla F\right\}
  \end{aligned}
\end{equation*}
This expression can be further simplified by making the shift $\hat F
\to \hat F+\frac 1 {1-F}$ to get
\begin{equation}
  \label{eqn:F}
  \begin{aligned}
    S&=\int \text{d}x \left[\rho \log\frac \rho F + (1-\rho) 
\log\frac {1-\rho}{1-F}\right]_0^T\\
    &+ \int \text{d}x
    \text{d}t \left\{ \hat F \dot F +\frac 1 2  \hat F^2 (\nabla F)^2 
       +\frac 1 2 \nabla \hat F \nabla F \right\} \\
  \end{aligned}
\end{equation}
The overall mapping from the initial $(\rho, \hat \rho)$ to $(F,\hat F)$ reads
\begin{equation}
\label{eqn:comp}
  F= \frac{\rho}{\rho + (1-\rho)e^{\hat \rho}};\;\; \hat F= (1-\rho)(e^{\hat \rho}-1)-\rho (e^{-\hat \rho}-1)
\end{equation}
The boundary conditions (\ref{eqn:BC}) are now given by
$F(0,t)=\rho_0$, $F(L,t)=\rho_L$ and $\hat F(0,t)=\hat F(L,t)=0$ and
the equation of motions are
\begin{equation}
  \label{eqn:EQMF}
  \dot F = \frac 1 2 \Delta F - \hat F (\nabla F)^2;\quad \dot {\hat F}
=-\frac 1 2 \Delta \hat F - \nabla \left[ {\hat F}^2 \nabla F \right]
\end{equation}

Modulo an integration by part, the last integral in (\ref{eqn:F}) is of
the form $ \int \text{d}x \text{d}t \left\{ \hat F \dot F -
  {\cal{H}}_F \right\}$ where
\begin{equation*}
  {\cal H}=  - \hat F  {\frac {(\nabla F)} 2}^2
 \left[ \hat F - \frac{ \Delta F}{(\nabla F)^2}\right]= - \hat F {\frac{ (\nabla F)}2}^2 \left[  \hat F - 
    \frac{\delta V_F }{ \delta F(x)} \right]
\end{equation*}
This is very similar to (\ref{eqn:NLDB}), but with $V_F \equiv \int \text{d}x
\; \ln (\nabla F)$. Quite surprisingly, we have once again obtained an
action formally deriving from a stochastic dynamics, satisfying
the detailed balance symmetry induced by
\begin{equation}
  \label{eqn:NLDBworks}
  \hat F \to \hat F + \frac{\Delta F}{(\nabla F)^2}\qquad (\hat
  F,t)\to(-\hat F,T-t)
\end{equation}
As in all the previous examples, two classes of solutions are
immediately available. First, $\hat F=0,\, \dot F = \frac 1 2 \Delta
F$ corresponds to a downhill diffusive solution, equivalent to $\dot
\rho = \frac 1 2 \Delta \rho$ and which is not what we are looking
for. Instead, an uphill trajectory is provided by requiring $\hat F
=\frac{\delta V_F }{ \delta F(x)}$. Together with the equation of
motion (\ref{eqn:EQMF}), it implies
\begin{equation}
\label{eqn:instanton}
  \dot F = - \frac 1 2 \Delta F;\qquad \rho = F + F (1-F) \frac{\Delta F}{\nabla F^2}
\end{equation}
Amazingly, this time, {\em its solution satisfies both spatial and
  temporal boundary conditions}, as is easy to check. The
corresponding action is the large deviation function
\begin{equation}
  \label{eqn:LDF}
  {\cal F}=\!\!\int\!\! \text{d}x \left[\rho \log\frac \rho F + (1-\rho) 
    \log\frac {1-\rho}{1-F}+\log \nabla F\right]_0^T
\end{equation}
\begin{center}
{\bf Dual Model}
\end{center}
The above derivation consists of changes of variables that read like a
sequence of miracles, at the end of which we are able to find an
`uphill' solution that satisfies the boundary conditions.  In the
following we show that all the surprises can be though of as deriving
from only one: the action for a chain in contact with two reservoirs
can be mapped, at the level of large deviations, to that of an open,
isolated chain. Starting from the action (\ref{eqn:F}), we introduce
the non-local variables
\begin{equation}
\label{eqn:primedef}
\hat F'=  \nabla F;
\qquad \hat F = \nabla\left[F'-\frac 1 {\hat F'}\right]
              = \nabla F' + \frac {\nabla \hat F'}{\hat {F'}^2}
\end{equation}
which takes the action into
\begin{equation*}
  \label{eqn:Fprime}
  \begin{aligned}
     \!\!&S\!=\!\!\int\!\!\text{d}x\!\left[ \rho \log\frac \rho F +
      (1-\rho) \log\frac
      {1-\rho}{1-F}+\log \hat F' -F' \hat F'\right]_0^T\\
    +& \int \text{d}x \text{d}t \left\{ \hat {F'} \dot F' +\frac
      1 2 \hat {F'}^2 (\nabla F')^2 +\frac 1 2 \nabla F' \nabla \hat
      F'\right\}
  \end{aligned}
\end{equation*}
Remarkably, the form of this action is, up to boundary terms,
the same as (\ref{eqn:F}). This suggests that we complete the mapping
\begin{equation}
  \label{eqn:change}
  (\rho,\hat \rho)\longrightarrow (F,\hat F)
 \overset{\text{non-local}}{\longrightarrow} (F',\hat F')
 \longrightarrow (\rho',\hat \rho')
\end{equation}
where the relation between $(\rho',\hat \rho')$ and $(F',\hat F')$ is
of the same form as (\ref{eqn:comp}). One then obtains
\begin{equation}
  \label{eqn:mapaction}
  \begin{aligned}
    &S\!=\!\!\!\int\!\!\text{d}x\!\left[\rho
 \log\frac \rho F + (1-\rho) \log\frac
    {1-\rho}{1-F}+\log \nabla F\right.\\
    -&\left. \frac{\rho'-F'}{1-F'}-\rho' \log\frac
 {\rho'} {F'} - (1-\rho') \log\frac
    {1-\rho'}{1-F'}\right]_0^T\!\!\!+S'
  \end{aligned}
\end{equation}
where $S'=\int \text{d}t \text{d}x \left\{ \hat \rho' \dot \rho ' -
  {\cal H}'\right\}$ and $ {\cal H}'$ is formally equivalent to 
(\ref{eqn:hamdens}) but for the primed variables:
\begin{equation}
  S'=\int \text{d}t \text{d}x \left\{ \hat \rho' \dot \rho ' - 
\frac 1 2 \sigma_{\rho'} \nabla \hat {\rho '}^2 + \frac 1 2 \nabla
 \rho ' \nabla \hat \rho'\right\}
\end{equation}
The overall change of variable (\ref{eqn:change}), which reads
\begin{equation}
  \label{eqn:nutshell}
\begin{aligned}
  \nabla \left[\frac 1 {1-e^{\hat \rho'}}\right]&=e^{\hat
    \rho}-1-\rho (e^{\hat \rho}+e^{-\hat \rho}-2)\\
  \nabla \left[ \frac{\rho}{\rho + (1-\rho) e^{\hat \rho}}
  \right]&=e^{\hat \rho'}-1-\rho' (e^{\hat\rho'}+e^{-\hat \rho'}-2)
\end{aligned}
\end{equation}
thus maps the action of the hydrodynamics limit of the SSEP into
another SSEP. We shall next show that the boundary conditions
transform in such a way that the dual chain is isolated. Before going
on, it is instructive to introduce the classical spins ${\bf S'}$
\begin{equation}
 \label{eqn:spin}
 S'_z= 2 \rho' -1\quad S'_+=  2 (1-\rho') e^{\hat \rho'}\quad 
S'_-=2 \rho' e^{-\hat \rho'}
\end{equation}
This is the usual connection between spin chains and particle models
\cite{Fogedby, Schutz} in the hydrodynamic limit. The Hamiltonian
density reads ${\cal H'}= -\frac 1 8 \nabla {\bf S'} \cdot \nabla {\bf
  S'}$~\footnote{The equations of motion read ${\bf \dot S}= \frac
  {\mathrm i} 2 {\bf \Delta S} \wedge {\bf S}$.}. It is invariant by
simultaneous rotation of all the spins, which means that to the three
'charges':
\begin{equation*}
2\rho'- 1=S'_z \quad; \quad   \hat F'=S'_z+iS'_y \quad; \quad
 \hat F'(1-2F')=S'_x-1
\label{charges}
\end{equation*}
correspond three currents
\begin{equation}
  \begin{aligned}
    &J_{\rho'}=-\frac 1 2 \nabla
    \rho' + \sigma_{\rho'} \nabla \hat \rho'  \;\;\;; \;\;\;  J_{\hat F'}=
    \frac 1 2 \nabla \hat F' + \hat F^{'2} \nabla F'\!\!\!\!\\
    &J_{\!\hat F'(1-2\hat F')}\!=\! (1-2F') \frac {\nabla \hat F'}2\!+\![\hat F'\!\!+\!{\hat {F'}}^{\!2}\!(1-2F')]\nabla F'\!\!\!\!
    \label{currents}
  \end{aligned}
\end{equation}
that are conserved in the bulk. The boundary conditions read for the
primed variables
\begin{equation}
  \label{boundaryprime}
  \nabla\left[F'-\frac 1 
    {\hat F'}\right]_{x=0,L}=0;\qquad \int_0^L \hat F'=\rho_l-\rho_0
\end{equation}
Note that all trajectories satisfying both the equations of motion
(\ref{eqn:EQMF}) and the boundary conditions satisfy also $\Delta F= \nabla
\hat F'=0$ on the boundaries. Together with the l.h.s. of
(\ref{boundaryprime}) and the definitions (\ref{eqn:comp}), this
implies that $\nabla F'$,$\nabla \hat F'$,$\nabla \rho'$ and $\nabla
\hat \rho'$ vanish separately at the boundaries, as do, consequently, all the
currents (\ref{currents}): {\em the dual model in the primed variables
  is an isolated chain}. This condition alone, supplemented with the
r.h.s.  of (\ref{boundaryprime}), encompasses all the original
boundary conditions.

Let us now turn to the trajectories. The stationary profile $\bar
\rho$ maps to a flat profile $\bar \rho'$, but whereas $\hat F'$ is
constrained by the r.h.s of (\ref{boundaryprime}), the precise value
of $\bar \rho'$ is arbitrary. {\bf Relaxations}: the diffusive
trajectories of the initial model correspond to $\dot \rho=\frac 1 2
\Delta \rho,\,\hat \rho=0$, and thus $\hat F=0$, as can be read in
(\ref{eqn:comp}). Together with (\ref{eqn:nutshell}), this implies
$\nabla \hat \rho'=0$. As the primed variables also satisfy the
equations of motion (\ref{eqn:hameq}), the resulting trajectories
evolve with $\dot \rho'=\frac 1 2 \Delta \rho'$. {\em Relaxations thus
  map into relaxations}. {\bf Excursions}: The instanton equations
(\ref{eqn:instanton}) imply $\nabla F'=0$
(cf. (\ref{eqn:primedef})). Using the relation (\ref{eqn:comp}) for
the primed variable, this yields $\nabla \hat \rho'-\frac{\nabla
  \rho'}{\sigma_{\rho'}}=0$. Together with the equations of motion
(\ref{eqn:hameq}), this shows that the densities evolve with 
$\dot \rho'=-\frac 1 2 \Delta \rho'$: {\em excursions are the time
  reverse of relaxations in the primed variables}. The action $S'[\hat
\rho',\rho']$ of such an uphill trajectory is $\int_0^L \text{d} x
[\rho' \log\rho' +(1-\rho')\log(1-\rho')]_0^T$.  As $F'$ is constant
along the instanton and $\int_0^L  \rho'$ is a constant of
motion, the overall action (\ref{eqn:mapaction}) reduces to the large
deviation function (\ref{eqn:LDF}) as it should.

Let us now show how the existence of the primed model accounts for the
many non-local symmetries of the initial one. The action of the
dual, isolated, chain is invariant under a composition of any
simultaneous rotation and reflection of all the spins, followed by
time-reversal. Written in the original variables, this corresponds to
a group of spatially non-local symmetries of the out-of-equilibrium
chain. Transformations of the primed model which do not conserve
$\int_0^L \hat F'=\frac{\rho_L-\rho_0}L$  correspond to
symmetries of the initial model broken by the boundary
conditions. This was for instance the case of (\ref{eqn:BrokNLDB}). If
we further impose the conservation of $\int_0^L \hat F$, we are left
with a transformation that in spin representation amounts to
$(S'_x,S'_y,S'_z,t)\to(-S'_x,S'_y,S'_z,T-t)$. This is in the primed
variables the non-local mapping (\ref{eqn:NLDBworks}) between
`downhill' diffusive solutions and the instantons of the initial
problem.
\begin{center}
{\bf Conclusion}
\end{center}
In this letter we have shown that the remarkable properties of the
hydrodynamic model allowing for its direct, explicit solution are
attributable to the existence of a dual, equilibrium model. Whereas
the derivation we have presented is very specific of the model we
studied, it can easily be extended to the other cases recently solved
within the Macroscopic fluctuation theory, including the Kipnis
Marchioro Presutti model~\cite{Bertini05}. A further extension to a
larger class of non-equilibrium systems or to microscopic models is
still an open and challenging question. The symmetry between
excursions and relaxations in the dual model is broken by the mixing
of variables due to the non-local mapping (\ref{eqn:primedef}). It
would thus be very interesting to see if one can construct non-local
quantities (like the currents of the dual model) which would be
symmetric, and measure them experimentally.

{\bf Acknowledgment} {We wish to thank T. Bodineau and B. Derrida for 
  interesting discussions}

\end{document}